# Time of life as it is in LiFeAs

A. A. Kordyuk[1,2], V. B. Zabolotnyy[1], D. V. Evtushinsky[1], T. K. Kim[1], I. V. Morozov[1,3], M. L. Kulić[4], R. Follath[5], G. Behr[1], B. Büchner[1], S. V. Borisenko[1]

[1]*Leibniz-Institute for Solid State Research, IFW-Dresden, D-01171 Dresden, Germany*
[2]*Institute of Metal Physics, 03142 Kyiv, Ukraine*
[3]*Moscow State University, Moscow 119991, Russia*
[4]*Institut for Theoretical Physics, Goethe University, Frankfurt am Main, Germany*
[5]*Helmholtz-Zentrum Berlin, BESSY, D-12489 Berlin, Germany*

**The time of life of fermionic quasiparticles, the distribution of which in the momentum-energy space can be measured by angle resolved photoemission (ARPES), is the first quantity to look for fingerprints of interaction responsible for the superconducting pairing[1]. Such an approach has been recently used for superconducting cuprates[2], but its direct application to pnictides was not possible due to essential three-dimensionality of the electronic band structure[3] and magnetic ordering[4]. Here, we report the investigation of the quasiparticle lifetime in LiFeAs, a non-magnetic stoichiometric superconductor[5-7] with a well separated two-dimensional band[8-10]. We have found two energy scales: the lower one contains clear fingerprints of optical phonon modes[11] while the higher scale indicates a presence of strong electron-electron interaction. The result suggests that LiFeAs is a phonon mediated superconductor with strongly enhanced electronic density of states at the Fermi level.**

In a search for the mechanism of superconducting pairing in pnictides, LiFeAs is, apparently, a key compound. Since it reveals the classical isotropic superconducting gap[10] and absence of any magnetic ordering, which may provide a basis for the spin-fluctuation alternative to phonon-mediated pairing as in case of cuprates[2], it is expected that the pairing glue is provided by phonons. On the other hand, the calculated strength of the electron-phonon coupling seems far from being sufficient to explain so high critical temperature ($T_c$ = 18 K)[11]. Therefore, the mission of the experiment is to clarify the actual value of the coupling strength and, if it is really insufficient, to uncover a missing piece of the high $T_c$ puzzle in LiFeAs.

Fortunately, from experimental point of view, LiFeAs provides a simplest case among the other pnictides to address those questions. First, it is a stoichiometric compound that exhibits superconductivity with a rather high transition temperature at ambient pressure without chemical doping[5-7], thus can be easily studied by the experimental techniques which require impurity clean samples. Second, it reveals a perfectly two-dimensional electronic band[8-10], well separated in momentum space from other bands, that gives possibility to precisely derive the quasiparticle self-energy[12] from ARPES spectra and analyse its structure[13-15]. Finally, LiFeAs cleaves between the two layers of Li atoms resulting in equivalent and neutral counterparts, offering a unique opportunity to overcome the problem of polar surface that can be crucial for the surface sensitive photoemission and tunnelling spectroscopies[16].

Figure 1a presents the Fermi surface map of LiFeAs, where the band of interest is contoured by the dashed line. The systematic ARPES study of this band[10] shows that it is really strongly two-dimensional: neither its position nor width varies with excitation energy and polarization. Figure 1 also shows the ARPES images, the cuts of the photoemission intensity in momentum-energy space, of two kinds: The spectra in panels b-e were measured with ultimate experimental resolution and statistics (see Methods), from which we extract the fine structure of band renormalization. Panel f shows the cut obtained by resampling the map data, from which we can estimate the behaviour of the self-energy on the energy scale of the band width.

Figure 2a shows the experimental dispersions, i.e. the positions of the maxima of the momentum distribution curves (MDC) as function of energy, for the cuts 1 and 2. Except a tiny effect of the superconducting gap in the close (~3 meV) vicinity of the Fermi level, the variation of band dispersion with temperature is negligible (the evolution of the whole quasiparticle spectrum can be seen in Fig. 1 b, c). Importantly, all the dispersions reveal three sharp kinks. Such kinks in the experimental dispersion are usually a consequence of interaction of the electrons with sharp modes residing at the kink energies in a bosonic spectrum[13-15]. All three kinks are clearly seen on the temperature integrated and smoothed dispersion curve (solid yellow curve), whose straight segments are highlighted by the dotted lines. Taking the 2nd derivative of this dispersion (dotted yellow curve in the inset) reveals the kink positions in a form of clear peaks. The energies of the observed kinks are remarkably close to the energies of the optical phonon modes (indicated by the vertical dashed lines): 15 meV, 30 meV, and 44 meV, recently calculated for this compound[11]. The lowest kink corresponds to the lowest phonon mode (121 cm$^{-1}$ $E_g$) and the highest kink corresponds to the highest mode (356 cm$^{-1}$ $A_{1g}$) while the middle kink fits to the energy of one of the intermediate phonon modes such as 240 cm$^{-1}$ $E_g$[11].

Now we turn to the quasiparticle scattering rate, or the imaginary part of the self-energy, which is inversely proportional to the quasiparticle lifetime. Observing such a clear fingerprint of the bosonic spectrum in the experimental dispersion, the fine structure of which is entirely associated with the real part of the self-energy, one should expect to see very similar structure in its imaginary part, Im$\Sigma$. In principle, the scattering rate is even more appropriate quantity to look for the details of the bosonic spectrum since their relation is much simpler (in the most trivial case of isotropic electron-boson coupling, constant electronic density of states, and zero temperature, the bosonic spectrum simply coincides with the differential scattering rate, $d$Im$\Sigma/d\omega$, see Methods). In Figure 2b, we show the MDC width, $\Delta k$, of the quasiparticle distribution, which is roughly proportional to the scattering rate (providing the Fermi velocity does not change much), along the same two cuts over the same energy range as in panel a. The first derivatives of integrated and normalized $\Delta k(\omega)$ dependences, shown in the inset, reveal the bosonic spectrum that is peaked at the same frequencies (slight displacements of the highest modes from 44 meV is a natural influence of the noise which highly increases with binding energy).

The contribution to the quasiparticle scattering rate of the phonon spectrum, which is confined below 44 meV, should saturate above that energy. This is exactly the case as one can see in Figure 2b. Since the real part of the self-energy rapidly decreases after the cut off of the phonon spectrum, we can neglect its contribution to the total renormalization soon after this energy and estimate the phonon coupling constant from the dispersion curves. The dotted and solid lines on the dispersion curve from cut 2 (Fig. 1a) represent the low energy experimental dispersion (that includes the renormalization by phonons) and the dispersion as it would be without coupling to phonons. Their velocities are $v_{low}$ = 0.24 eVÅ and $v_{high}$ = 0.35 eVÅ, respectively, and the value $\lambda_{eff}$ = $v_{high}/v_{low}$ − 1 = 0.46 gives an effective (i.e. renormalized by other non-phononic interactions) electron-phonon coupling strength[17].

Can the electron-phonon interaction, which we have determined, be responsible for the strong renormalization of the band width (about factor of 3) seen in LiFeAs[10]? Evidently not, since its contribution to the quasiparticle self-energy is confined within the energy range of phonons (44 meV). In order to understand the missing mechanism for the strong renormalization, in Figure 3, we show an overview plot of the scattering rate in LiFeAs on the scale of the band width and compare it to the same quantity measured along the nodal direction in a cuprate superconductor, the optimally doped $(Bi,Pb)_2Sr_2Ca_2CuO_{8-\delta}$ (BSCCO)[14]. Note that in both Fig. 2b and Fig. 3 (left axis) we show not strictly the scattering rate, but the MDC half-width, the value directly seen by ARPES. Its relation to the imaginary part of the self-energy and quasiparticle lifetime, $\tau$, is simple: $\hbar/\tau = -Im\Sigma = v_F \Delta k$, provided the Fermi velocity of the bare electrons, $v_F$, is known. Taking $v_F \approx 1$ eVÅ for LiFeAs[8-10], we add the energy scale as right axis to Fig. 3.

From the overview plot, one can see that the phonons are not the only contributors to the scattering rate. The phonon related step develops on a background of another contribution which shows linear dependence on energy both below (as indicated by the blue dashed line in Fig. 3) and above the phonon energy range and tends to saturate above 100 meV so that the total scattering rate reaches ~80 meV. The saturated value of the scattering rate can be estimated independently as the band broadening at its very bottom (see Fig. 1d). The half width of the energy distribution curve (EDC) at that momentum gives virtually the same value, which is shown in Fig. 3 by the circle (the horizontal error bars represent the uncertainty in Re$\Sigma$). This is independent evidence that 1 eVÅ is a good estimate for $v_F$, and, consequently, that the renormalization factor of 3 in LiFeAs is indeed a self-energy effect, caused mainly by the non-phonon scattering channel, which is natural to associate with an electron-electron scattering[14] with $\lambda_{el} = 2$. Then, the absolute value of the electron-phonon coupling strength can be estimated as $\lambda_{ph} = \lambda_{eff}(1 + \lambda_{el}) = 1.38$ (see Methods).

Having identified and evaluated the two constituents of the quasiparticle self-energy as the electron-phonon and electron-electron scatterings, we can reconsider the high $T_c$ problem in LiFeAs. There is an opinion, which is based on the calculated lower value $\lambda_{ph} = 0.29$[11] and on the McMillan's formula for $T_c$[18] modified by Allen and Dynes[19], that the phonon coupling in LiFeAs is by far not sufficient to explain the onset of superconductivity at 18 K. Direct substitution of the estimated value of $\lambda_{ph} = 1.38$ in the same formula increases $T_c$ from 0.005 K to 10 K, which is still lower than 18 K but already comparable.

Based on the extensive studies of the limitations of the McMillan's formula and its modifications[20-23], one can be confident that the remaining difference will be eliminated by taking into account the real spectrum of phonons[21], real electronic density of states (DOS)[22,23], and the intra-band electron-boson scattering[24]. While the evaluation of the contributions of each of those mechanisms in further increase of the $T_c$ waits for the numerical solution of the Eliashberg equation, this our study suggests that the electronic band structure itself plays an important role. The linear electron-electron scattering rate (see Fig. 3) is well known signature of an extended van Hove singularity (vHs) residing at the Fermi level[23]. The extended vHs is, in fact, observed as a flattening of the other band at the Fermi level in the center of the Brillouin zone (see Fig. 1d), as discussed in details in Ref. 10. The ability of the vHs to increase the critical temperature considerably, i.e. besides the simple influence of $\lambda$ by the effective DOS increase, is well known and has been discussed extensively for the cuprates[22,23].

Clearly, the main reason for much higher value of λ_ph, comparing to the theoretical estimates, is the increased electronic DOS due to strong renormalization. Since such a strong renormalization seems to be general for the pnictides[10,25], resolving the microscopic origin of this scattering can be a key to the problem of superconductivity in these compounds. Here we have concluded that the strong renormalization can be explained by the vHs at the Fermi level and strong electron-electron scattering, but have not discussed the origin of the latter.

Besides a straightforward Coulomb interaction, there is a possibility that the observed electron-electron scattering is dominated by electron coupling to the spin-fluctuations, which are generally expected to be strong in all the ferropnictides[26] and have been reported in LiFeAs by NMR study[27]. If the spin-fluctuations are really strong in LiFeAs, they can help to increase the $T_c$ either indirectly, contributing to the DOS renormalization, or providing another, non-phonon mechanism for the superconducting pairing[26]. While the adequate consideration of the role of the spin-fluctuations in LiFeAs requires a clear result from the inelastic neutron scattering experiment, the present ARPES study does not provide evidence for their role in the formation of the fermionic spectrum, as in case of cuprates[2]. In particular, any additional renormalization below $T_c$ (apart from the opening of the superconducting gap and reducing of the temperature broadening, as shown in Fig. 1 b and c) is virtually absent in LiFeAs, while it is routinely observed for the antinodal fermions in cuprates and known as the coupling to the spin-1 resonance[28]. On the other hand, the comparison of the scattering rates of the quasiparticles from the two-dimentional band of LiFeAs and from the nodal region of BSCCO (see Fig. 3) shows that the phonon contribution is much stronger in the former, even assuming that the step-like behavior in the latter is also related to phonons[29].

We acknowledge discussions with I. Mazin, A. Yaresko, S.-L. Drechsler, T. Dahm, I. Eremin, and E. Pashitskii. The project was supported by the DFG under Grants No. KN393/4, BO 1912/2-1, BE1749/13, 486RUS 113/982/0-1, and priority programme SPP 1458. I.V. Morozov also acknowledges support from the Ministry of Science and Education of the Russian Federation under State Contract P-279.

**Methods**

**ARPES experiments** have been performed at the "1^3" beamline at BESSY equipped with SES 4000 analyser and $^3$He cryo-manipulator with the lowest temperature on the sample below 0.9 K. The spectra for the analysis of the dispersion and scattering time anomalies have been measured along several cuts (cuts 1 and 2 in the paper) the position of which in the reciprocal space and excitation energy have been optimized for the maximum of photoemission intensity at 4 meV and 0.01 Å$^{-1}$ of the energy and momentum resolution, respectively. The cut 3 has been resampled from the map. The experimental statistics in this cut (~40 counts per pixel) is 20-40 times lower but still sufficient to derive the MDC width behaviour up to about 100 meV and determine the EDC width at the bottom of the band.

**Data treatment.** For a momentum independent phonon (generally, boson) spectrum, $F(\Omega)$, and electron-phonon coupling, α, the phononic contribution to the scattering rate (or the imaginary part of the quasiparticle self-energy) can be written as[2]

$$\text{Im}\Sigma(\omega) \propto \int_{-\infty}^{\infty} \alpha^2 F(\Omega)[n(\Omega) + f(\omega - \Omega)]\text{DOS}(\omega - \Omega)d\Omega,$$

where $n$ and $f$ stand for the Bose and Fermi functions, respectively. For constant DOS and zero temperature, the first derivative of $d\text{Im}\Sigma(\omega)/d\omega \propto \alpha^2 F(\omega)$. Until the width of the peaks in $\alpha^2 F$ remains much smaller than the width of DOS anomalies, their positions should not deviate noticeably from the positions of the peaks in $d\text{Im}\Sigma(\omega)/d\omega$. The real and imaginary parts of $\Sigma$ are related by the Kramer-Kronig transform (see Ref. 12 for the examples of $\Sigma$ parts in cuprates). This generally means that in order to derive $\alpha^2 F$ from $\text{Re}\Sigma$ one needs to solve an Eliashberg-kind equation[1,13]. Nevertheless, for the presented accuracy, the corresponding peak positions in $d\text{Im}\Sigma(\omega)/d\omega$ and $d^2\text{Re}\Sigma(\omega)/d\omega^2$ coincide (see Refs. 12, 13). So, the second derivative of the MDC dispersion, though not revealing the true profile of $\alpha^2 F(\Omega)$, still gives a good estimate for the positions of the peaks in the bosonic spectrum.

Neglecting the phonon contribution to $\text{Re}\Sigma(\omega)$ essentially above the phonon cut-off frequency, one can estimate the electron-phonon coupling strength from the experimental dispersion, $\varepsilon(k)$, provided that two linear regions, for which $d\varepsilon(k)/dk = \varepsilon(k)/k = v$, are observed. The low- and high-energy velocities defined for those regions can be expressed then through the bare Fermi velocity, $v_0$, and the electron-phonon and electron-electron coupling constants as follows: $v_{\text{low}} = v_0/(1 + \lambda_{\text{ph}} + \lambda_{\text{el}})$, $v_{\text{high}} = v_0/(1 + \lambda_{\text{el}})$. Then $\lambda_{\text{ph}} = \lambda_{\text{eff}}(1 + \lambda_{\text{el}})$, where $\lambda_{\text{eff}} = v_{\text{high}}/v_{\text{low}} - 1$, is an effective, i.e. renormalized by an electron-electron interaction, electron-phonon coupling strength which is usually measured in ARPES experiment[17].

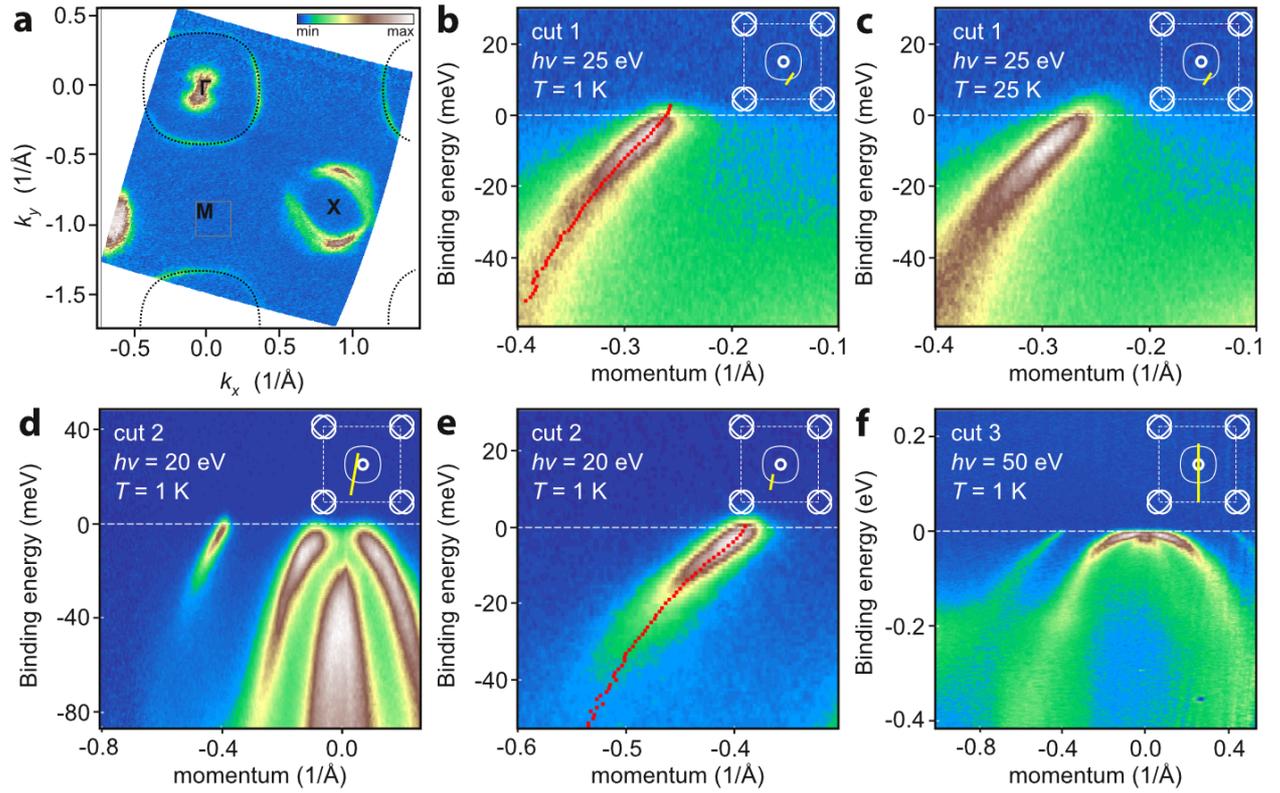

**Fig. 1. Quasiparticle spectrum of LiFeAs as seen by ARPES. a**, The Fermi surface (FS) map. The narrowest contour around Γ-point (indicated by dotted line) corresponds to the hole-like FS sheet which is formed by the two-dimensional band of interest. **b-f**, Different cuts of this band reveal its behaviour as function of binding energy and in-plane momentum. The position of each cut is shown on the FS sketch in the inset on each panel. Red squares on panels **b** and **e** trace the experimental dispersion.

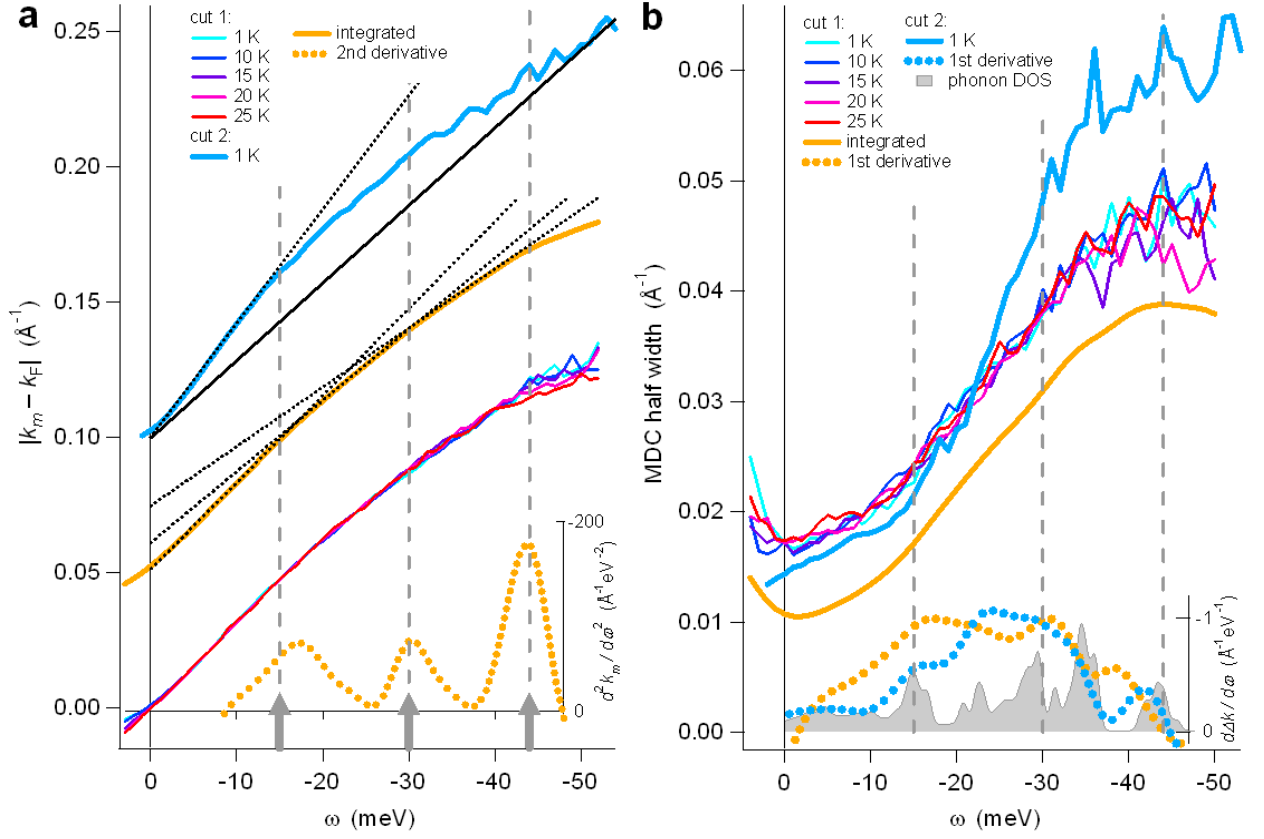

**Fig. 2. Fingerprints of low energy interaction. a**, Experimental (or renormalized) dispersions: the positions of MDC maxima, $k_m$, as function of energy, $\omega$, for cut 1 at different temperature, for the same cut but temperature integrated and smoothed, and for cut 2. The last two curves are shifted up by 0.05 and 0.1 Å$^{-1}$, respectively. The inset shows the 2nd derivative of the dispersion *vs* the same energy axis. **b**, MDC width (the half width at the half maximum) which is proportional to the inversed lifetime of the fermionic quasiparticles. The similarly integrated curve is shifted down. The inset shows the differential scattering rates for cuts 1 and 2 on top of the phonon density of states estimated from Ref. 11 in a false grey scale. On both panels, the vertical dashed lines indicate the positions of three optical phonon modes: 15 meV (the lowest one), 30 meV, and 44 meV (the highest one).

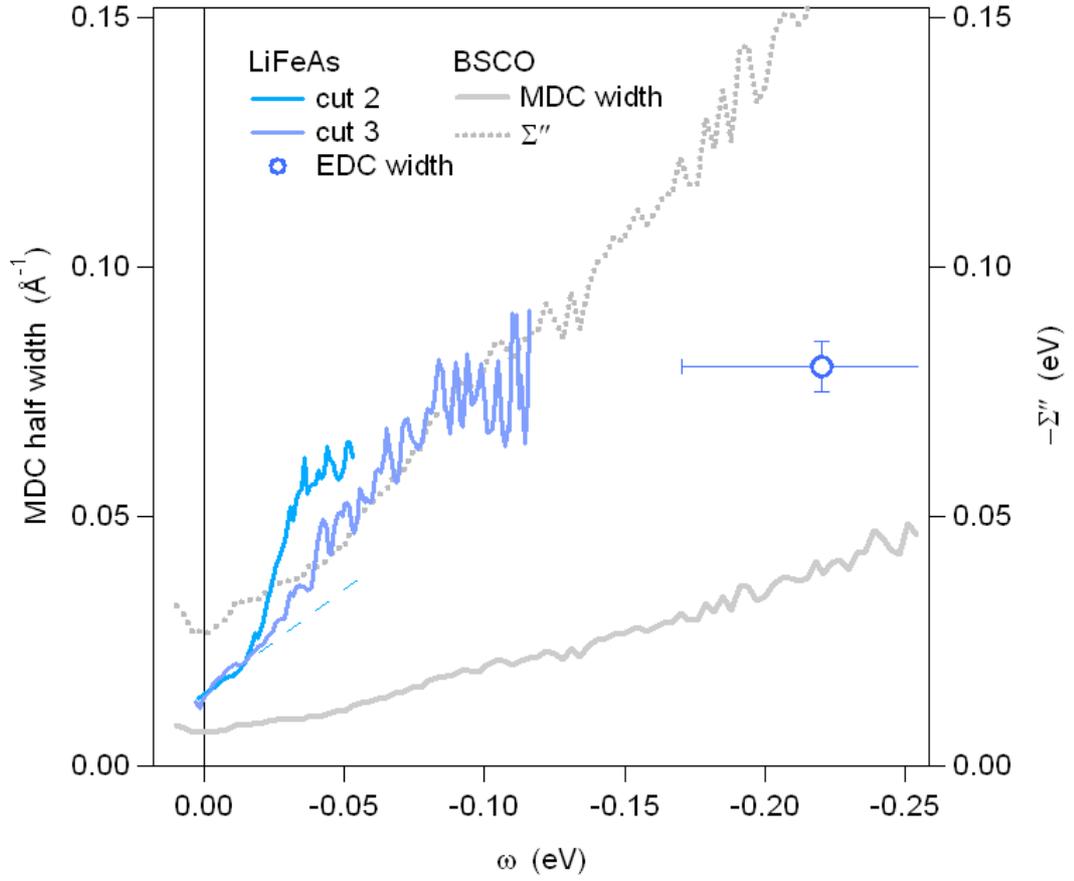

**Fig. 3. Scattering rate on large energy scale.** MDC width and the imaginary part of fermionic self-energy (the scattering rate) for the cuts 1 and 2. MDC width for $(Bi,Pb)_2Sr_2Ca_2CuO_{8-d}$ (BSCCO)[14] is shown for comparison. The EDC width at the bottom of the band and the imaginary part of the self-energy for BSCCCO refer to the right axis.